# Ultrastrong light–matter coupling in near-field coupled split-ring resonators revealed by photocurrent spectroscopy

Jing Huang[1], Jinkwan Kwoen[2], Yasuhiko Arakawa[2], Kazuhiko Hirakawa[1], and Kazuyuki Kuroyama[1],

[1]*Institute of Industrial Science, University of Tokyo, 4-6-1 Komaba, Meguro-ku, Tokyo 153-8505, Japan*

[2]*Institute for Nano Quantum Information Electronics, University of Tokyo, 4-6-1 Komaba, Meguro-ku, Tokyo 153-8505, Japan*

**Abstract**

Landau polaritons arising from the coupling between cyclotron resonance and terahertz split-ring resonators (SRRs) have served as a central platform for exploring ultrastrong light–matter interaction for more than a decade. Over this period, a wide variety of SRR architectures, differing in size, geometry, and even material composition, have been investigated. However, the regime of near-field coupled SRRs has remained largely unexplored. Here, we demonstrate ultrastrong coupling using photocurrent spectroscopy in two prototypical near-field configurations: a SRR dimer and a topological SRR chain. The measurements reveal hybridization not only with bright resonant modes but also with optically dark modes and topological edge modes, highlighting the exceptional sensitivity of the photocurrent spectroscopy. Moreover, the engineered near-field interactions allow the study of multi-mode ultrastrong coupling and the interplay between topological band structure and cavity quantum electrodynamics.

# I. INTRODUCTION

The interaction between light and matter in the regime where their coupling strength $\Omega$ exceeds 10% of the bare transition frequency $\omega$—the ultrastrong coupling (USC) regime—has attracted intense theoretical and experimental interest across photonics and condensed matter systems. In this regime, the conventional rotating-wave approximation breaks down, and new hybridized eigenstates, known as polaritons, exhibit nontrivial quantum properties such as ground-state virtual excitations [1] and nonperturbative light–matter correlations [2-4]. Recently, the experimental realizations of USC have opened up new possibilities for controlling quantum states of matter [5-9] and generating squeezed light in frequency domain [10].

A powerful platform for realizing USC in the terahertz range is Landau polariton [11], which is formed by coupling cyclotron resonance (CR) of a two-dimensional electron gas (2DEG) to a terahertz cavity. The CR frequency, $\omega_{CR} = eB/m^*$, can be continuously tunable by the magnetic field $B$, allowing precise control of the resonance condition with the cavity mode. When CR couples to terahertz split-ring resonators (SRRs), the combination of the SRR's subwavelength field confinement and the large dipole moment of the CR leads to exceptionally strong coupling [11, 12]. To date, an extreme normalized coupling strength of up to $\Omega/\omega_{SRR} = 3.19$ [13], has been achieved in Landau polariton systems—the highest reported among all light–matter coupled platforms.

A wide range of metallic resonator geometries have been investigated, including complementary SRRs [14], dipole-shaped SRRs designed to reduce the number of coupled electrons [15], nano-scale SRR to study polaritonic non-locality [16] and even superconducting SRRs that suppress dissipative losses [17, 18]. However, near-field coupled SRR architectures, such as SRR dimers, topological SRR chains and arrays, have received comparatively little attention in this context. These systems exhibit rich photonic properties, including highly tunable multimode spectra [19, 20], which can enhance ground-state virtual matter excitations, and topologically protected states [21, 22, 23] that enable the realization of ultrastrongly coupled topological polaritons. Nonetheless, their experimental study has been hindered by the limitations of conventional optical transmission/absorption spectroscopy, which is sensitive only to bright modes of resonators, whereas many near-field coupled SRR modes are optically dark [24, 25] Moreover, the far-field optical spectroscopy lacks the spatial resolution required to distinguish bulk modes from topological edge modes.

Recently, photocurrent spectroscopy has emerged as a powerful alternative [26, 27, 28], exploiting quantum Hall edge channels as local probes for light-matter interaction. This approach has already demonstrated high sensitivity in detecting USC in single meta-atom [26, 27] and in observing strong coupling between few-electron quantum dots and terahertz cavities [28].

Here, we employ photocurrent spectroscopy to investigate the USC between the CR of a high-mobility 2DEG and various near-field coupled SRR architectures—first a SRR dimer exhibiting multimode resonances, and then a topological SRR chain supporting spatially localized edge states. By comparing the simulated optical absorption spectra with the experimentally measured photocurrent spectra, we demonstrate the consistency between the two approaches and highlight the capability of photocurrent spectroscopy, which provides a spatially resolved probe of both bright and dark polaritonic modes supported by near-field coupled SRR structures. Notably, the USC realized in the topological SRR chain constitutes a topological polariton in the far-infrared regime, extending previous realizations in the visible-light domain [29].

## II. Results

**Photocurrent spectroscopy of ultrastrong coupling in a SRR dimer**

We first studied the USC in a SRR dimer, which consists of two SRRs placed in close proximity to enable strong near-field interaction. Figure 1(a) shows the SEM image of the dimer structure: a pair of identical SRRs are separated by 2 μm, fabricated on top of the GaAs surface. Two fine gate extensions are also defined to further enhance the local electric field. As the separation of the two SRRs is much smaller than the resonance wavelength of the SRRs, strong cavity-cavity interaction between the two SRRs is achieved and the resonance peak is split into two distinct eigenmodes [19, 20]. The resonance behavior of the SRR dimer is calculated using FE simulations (see Sec. 1 in Supplementary Materials). Figure 1(b) shows the amplitude of the in-plane electric field in one of the SRR gaps as a function of excitation frequency, $f_{THz}$. The SRR dimer shows two cavity modes: one peak appears at $\omega_1 = 0.94$ THz and the other at $\omega_2 = 0.76$ THz, with their electric field distributions shown in the top and bottom panels of Figs. 1(c), respectively. We refer to the resonant mode at higher (lower) frequency as the symmetric (antisymmetric) mode, since the resonant polarizations in the two SRR gaps are oriented in parallel (opposite) directions (see Sec. 2 of Supplementary Materials).

We then simulated the optical absorption spectra of the coupled system of the SRR dimer and 2D electrons under various perpendicular magnetic fields (see Sec. 3 in Supplementary Materials), as shown in Fig. 1(d). While a linear dispersion corresponds to the cyclotron resonance of the uncoupled 2DEG, only one avoid-crossing feature with two polariton branches appears when the cyclotron frequency, $\omega_{CR}$, is on resonance with the symmetric mode of SRR dimer (i.e., $\omega_1 = \omega_{CR}$). No clear absorption of the antisymmetric mode appears because it possesses only an electric quadrupole moment, which cannot be coupled efficiently to the far-field excitation [24, 25]. Using a fitting procedure based on the Hopfield model [11] for a single photon mode and a single matter mode (referred to as the single-mode Hopfield model in the following), we fit the simulated absorption spectra and extract a normalized coupling ratio of $\Omega_1^{sim}/\omega_1 = 0.17$, as shown by the red fitting curves in Fig. 1(d). This value indicates that the light-matter interaction lies in the USC regime. Here, $\Omega^{sim}$ represents the light–matter coupling strength obtained from the FE-simulation, will later be compared with the experimentally determined coupling strength $\Omega^{exp}$ from photocurrent spectroscopy.

Photocurrent spectroscopy was performed using a standard lock-in technique in a $^3$He cryostat with a superconducting magnet at a temperature of 0.34 K. The sample is placed on a Si hemispherical lens to focus the incident monochromatic terahertz radiation. The detection mechanism of the photocurrent spectroscopy is based on nonequilibrium electron distributions among multiple quantum Hall edge channels, which have been investigated in our previous studies [26, 27]. (More details are provided in Sec. 4 of Supplementary Materials). Figures 2(a) and 2(b) show photocurrent spectra of the SRR dimer sample at zero source-drain bias ($V_{SD} = 0$), when a negative gate voltage is applied to either the top- or bottom-SRR, respectively. In contrast to the absorption spectra, the photocurrent spectra reveal two distinct anti-crossing features, with the lower polariton arising from the coupling to the antisymmetric mode clearly resolved. The appearance of the antisymmetric mode in the photocurrent spectra can be attributed to two factors:

1. the antisymmetric mode can be excited because a single SRR directly senses the spatially inhomogeneous terahertz field, whereas in a conventional SRR array, the field inhomogeneity is

averaged out; and

2. the quantum Hall edge channels serve as local probes that are sensitive to near-field distributions, allowing the antisymmetric mode to be detected through the photocurrent response.

Because the symmetric and antisymmetric modes are orthogonal over the full space [30], their electric field distributions follow the orthogonality condition:

$$\int_V \boldsymbol{E}_1(\boldsymbol{r})\boldsymbol{E}_2^*(\boldsymbol{r})d^3r = 0, \tag{2}$$

where $\boldsymbol{E}_1(\boldsymbol{r})$ and $\boldsymbol{E}_2(\boldsymbol{r})$ denote the electric field distributions of the symmetric and antisymmetric modes. Consequently, their couplings to the CR can be treated independently. In line with previous studies [12], we first fit the photocurrent spectra using two separate Hopfield models:

$$\mathcal{H} = \sum_{\nu=1,2}\left[\hbar\omega_\nu a_\nu^\dagger a_\nu + \hbar\omega_{CR}b_\nu^\dagger b_\nu + i\hbar\Omega_\nu(a_\nu^\dagger + a_\nu)(b_\nu^\dagger - b_\nu) + \hbar\frac{\Omega_\nu^2}{\omega_{CR}}(a_\nu^\dagger + a_\nu)^2\right], \tag{3}$$

where $a_\nu$ ($a_\nu^\dagger$) and $b_\mu$ ($b_\mu^\dagger$) are the annihilation (creation) operators for cavity photon mode $\nu$ and matter excitation mode $\mu$, respectivrly. We use $\omega_1 = 0.94$ THz and $\omega_2 = 0.76$ THz are the calculated cold cavity mode frequencies used as input parameters. The resulting polariton dispersions are plotted as red and magenta curves in Figs. 2(a) and 2(b). From these fits, we extract normalized coupling strengths $\Omega_1^{\text{exp}}/\omega_1 = 0.15$ and $\Omega_2^{\text{exp}}/\omega_2 = 0.17$, both in the USC regime and in close agreement with the estimated value $\Omega_1^{\text{sim}}$ obtained from the absorption simulation.

In the CR-SRR dimer coupled system, although the single-mode Hopfield models can reproduce the main anti-crossing features in both simulated absorption and measured photocurrent spectra, a closer inspection of the spectra reveals a mismatch between the fitted polariton branches and the data, especially for the upper polariton branch in the optical absorption spectra in Fig. 1(d). These deviations are caused by the finite spatial overlap of the two cavity modes [30, 31, 32], which cannot not be fully captured by independent single-mode fits.

The spatial overlap of the symmetric and antisymmetric modes originates from the breakdown of orthogonality condition Eq. (2) when the integration is restricted to a sub-domain, such as the 2DEG plane, rather than the full three-dimensional space [30, 31, 32]. Following an elegant theory from Ref. [30], one can define a so-called overlap coefficient $\eta_{12}$ that quantifies the spatial overlap between two modes in the 2DEG plane. Here, $\eta_{12}$ ranges from 0 to 1, corresponding to completely isolated and fully overlapping modes, respectively. (More details of the spatial mode overlap and the multi-mode analysis for the SRR dimer are provided in Secs. 5 and 6 of the Supplementary Materials). In the CR-SRR dimer coupled system, we calculate the overlap coefficient $\eta_{12}$ between symmetric and antisymmetric modes using FE simulation, yielding $\eta_{12} = 0.19$, which corresponds to an intermediate overlap regime [32]. Incorporating this overlap into a multi-mode Hopfield model yields a significantly improved agreement with the spectra, as shown in Figs. S2 and S3 of the Supplementary Materials. Although the mode overlap in the present CR-SRR dimer system is relatively small, which prevents the clear appearance of the characteristic "S-shaped" dispersion typically associated with multi-mode USC [30, 31, 32], this limitation could, in principle, be alleviated through spatial structuring of the matter excitation [31].

Lastly, we emphasize that the photocurrent polarity depends not only on the magnetic field but also on which SRR is gated, reflecting its chiral nature, as illustrated in Figs. 2(a) and 2(b). This SRR-dependent chirality originates from the selective interaction of the quantum Hall edge channels with the cavity field. As shown in Figs. 2(c) and 2(d), when the top (bottom) gate is activated, the top (bottom) edge channels enter the field-enhancement region, generating a non-equilibrium electron distribution along the corresponding edge channel and producing the photocurrent. Within the Landauer-Büttiker formalism, at $V_{SD} = 0$, the photocurrent in the SRR dimer system at integer filling factor $v = n$ can be expressed as

$$\Delta I_{\text{THz}}^{v=n}(B) = sgn(B) \cdot \gamma_{geo} \cdot \frac{2e}{h} \Delta\mu_{n-1}, \tag{4}$$

where $\Delta\mu_{n-1}$ denotes the effective change in electrochemical potential resulting from polariton excitation [26, 27] and decay [33, 34] between the n[th] and n-1[th] Landau levels. The geometrical factor $\gamma_{geo} = \pm 1$ labels the propagation direction of the edge channel interacting with the cavity photon field and is selected by activation of the top ($+1$) or bottom ($-1$) SRR gate. This relation supports our model of non-equilibrium electron distribution and highlights the local probing capability of the quantum Hall edge channels, which is sensitive to spatially confined light–matter interactions.

**Spatially-resolved ultrastrong coupling in a topological SRR chain**

We then extend our structure into a topological SRR chain, as shown in Fig. 3(a). In this device, one unit cell consists of two SRRs (see the two leftmost SRRs enclosed by blue dashed rectangle in Fig. 3(a)), and two such unit cells form the SRR chain. According to the Su–Schrieffer–Heeger (SSH) model, when the intracell coupling is weaker than the intercell coupling, a topological edge state emerges within the band gap between the antisymmetric and symmetric bulk states [21]. From our FE simulation, we confirmed that the ratio of the intracell coupling to the intercell coupling is 0.38, which less than one, indicating that the SRR chain is in the topological phase. (See Sec. 7 in Supplementary Materials for details on the intracell coupling between SRRs).

The resonant behavior of the topological SRR chain is confirmed by FE simulation when probing the in-plane electric field spectra in different SRR gaps, as shown in Fig. 3(b). When probing the bulk SRRs, two resonant peaks appear at and $\omega_{\text{sym}} = 0.96$ THz and $\omega_{\text{anti}} = 0.74$ THz, analogous to the SRR dimer. The enlarged mode splitting compared with the SRR dimer originates from the intracell inductive coupling provided by the two edge SRRs. In contrast, when probing the edge SRRs, only a single dominant peak at $\omega_{\text{edge}} = 0.88$ THz is observed, corresponding to the topological edge state. Figures 3(c) and 3(d) display the electric field distribution of the bulk modes and the edge mode, respectively. While the symmetric and antisymmetric modes exhibit strong field enhancement within the bulk SRRs, the topological edge mode shows a more localized field enhancement confined to the one edge SRR and extending weakly toward the opposite edge. Such spatial profile is consistent with the estimated localization length derived from the ratio between the intracell and the intercell coupling, which is approximately 100 μm—larger than the total chain length of 80 μm—explaining the observed extension of the edge state across the structure.

Figure 4(b) shows the simulated optical absorption spectrum of the topological SRR chain coupled to the 2DEG. Consistent with the simulation of the coupled CR-SRR dimer system, only the bright polariton branches originating from the bulk symmetric mode and the bright edge mode can be observed. However, because these two modes are closely spaced in frequency and the SRR cavity has a low quality factor, the resulting spectral features exhibit significant overlap. The spectra are fitted using two independent single-mode Hopfield models, as indicated by the red curves in Fig. 4(b), giving normalized coupling strengths of $\Omega_{sym}^{sim}/\omega_{sym} = 0.15$ for the bulk symmetric mode (red solid curves) and $\Omega_{edge}^{sim}/\omega_{edge} = 0.15$ for the topological edge mode (red dashed curves). It is important to note that, despite the different spatial distributions of electric-field enhancement, optical absorption spectroscopy measures only the collective response of the SRR chain, and therefore cannot probe the bulk symmetric and bright edge modes independently.

We now analyze the photocurrent spectra measured on the SRR chain. Figure 4(a) shows a simplified schematic of the gating configuration used in the measurement. When photocurrent spectra are measured with one of the bulk SRRs gated, four polariton branches appear, indicating USC to both the antisymmetric and symmetric bulk modes, similar to the case of the SRR dimer, as shown in Fig. 4(c). In contrast, when the edge SRR is gated, only a single anti-crossing corresponding to USC with the topological edge state is observed, as shown in Fig. 4(d). Fitting these photocurrent spectra using independent single-mode Hopfield models yields normalized coupling strengths of $\Omega_{sym}^{exp}/\omega_{sym} = 0.14$, $\Omega_{anti}^{exp}/\omega_{anti} = 0.16$, and $\Omega_{edge}^{exp}/\omega_{edge} = 0.15$, all in the USC regime (see red and magenta curves in Figs. 4(c) and 4(d)) and in good agreement with the values inferred from the optical absorption simulations.

These results highlight the spatial selectivity of photocurrent spectroscopy, which enables the independent detection of USC for spatially separated polaritonic modes. This capability provides a powerful means to investigate USC phenomena in more complex SRR architectures, such as topological SRR arrays [23] or multilayer SRR structures for an enhanced Q-factor [35].

**III. Discussion**

We have demonstrated that photocurrent spectroscopy serves as a probe of ultrastrongly coupled Landau polariton in near-field coupled SRR architectures. By exploiting quantum Hall edge channels as sensitive local detectors, this technique enables direct access to both bright and dark polaritonic modes, which are often inaccessible to conventional far-field optical spectroscopy. The local nature of detection allows mapping of the light-matter interaction in different parts of the coupled SRR systems. We applied photocurrent spectroscopy to near-field coupled SRR architectures, including SRR dimers and topological SRR chains. In the SRR dimer, we resolved USC between CR and both the bright and dark modes, and we discussed how mode overlap modifies the resulting energy dispersion. Extending this concept to a topological SRR chain enabled spatially resolved probing of bulk and edge USC, where the edge-state polaritons were selectively detected by gating the edge SRRs.

Overall, the measurements demonstrate the potentials of near-field coupled SRRs: their ability to engineer and control multimode or spatially localized electromagnetic fields, and their compatibility with electrical readout via photocurrent spectroscopy. Together, these properties make near-field coupled SRR platforms highly attractive for exploring multimode light-matter phenomena, such as the realization of superstrong coupling [36] and the breakdown of single mode approximation [37, 38], and for integrating topological physics into solid-state cavity quantum electrodynamics, such as in the realization of topological polaritons. The topological SRR chain, whose edge states are robust against defects and impurities [21], could serve as loss-resistant channels for quantum-state transfer between spatially separated quantum emitters, such as quantum dots coupled to the chain ends, potentially benefiting the scalability of semiconductor-based quantum computers [39] and the development of topological polariton lasing in terahertz regime [29].

**IV. Online Methods**

**Sample preparation**

The samples were fabricated on modulation-doped GaAs/AlGaAs heterojunctions. The single SRR and SRR dimer devices were fabricated on a wafer with a carrier density of $1.86 \times 10^{11}$ cm$^{-2}$ and a mobility of $0.87 \times 10^6$ cm$^2$/Vs at 0.3 K. The topological SRR chain was fabricated on a different wafer, which exhibited a carrier density of $2.25 \times 10^{11}$ cm$^{-2}$ and a mobility of $1.30 \times 10^6$ cm$^2$/Vs. The SRRs (fine gates) were fabricated using an Electron beam lithography with *JEOL JBX-6300SG* followed by deposition of 10 nm titanium and 150 (30) nm of gold and a lift-off process. The bonding pad and the mesa were fabricated using a direct laser writing lithography with *Heidelberg µPG 101*. The resonators and the CR-resonator coupled system were simulated using *CST microwave studio* to confirm their resonant behaviors and optical transmission spectra.

To perform photocurrent measurements, the devices were then glued with varnish on a hemispherical silicon lens to tightly focus the terahertz radiation onto the SRR.

**Terahertz photocurrent spectroscopy**

We carry out all the measurements in a $^3$He cryostat with base temperature equal to 0.34 K when a terahertz window is opened. A simplified schematic of the photocurrent measurement setup is provided in Sec. 8 in the Supplementary Materials. Terahertz radiation is introduced through a window opened at the bottom of the cryostat. To suppress blackbody radiation, two multi-mesh low-pass filters with cut-off frequencies of 10 THz (50 K stage) and 6 THz (4 K stage) are installed. Additionally, black polypropylene sheets, which are opaque in the visible but transparent in the terahertz range, are placed in the optical path to block unwanted visible light.

A monochromatic THz beam from a uni-traveling-carrier photodiode, electrically modulated at a fixed frequency, passes through the THz low-pass filters and is focused onto the device using a metallic waveguide and a silicon lens. The THz-induced photocurrent is detected using a standard lock-in amplifier (*Stanford Research Systems SR830*).

**ACKNOWLEDGEMENTS**

This work was supported by JSPS International Joint Research Program (JPJSJRP20221202), KAKENHI from JSPS (JP25K00012, JP25K01691, JP25K01694, JP24K21526, JP22K03481, JP20K14384, JP20H05660, JP15H05700), JST, PRESTO Grant No. JPMJPR2255, Research Foundation for Opto-Science and Technology, and Murata Science and Education Foundation.

**FIGURE CAPTIONS**

FIG. 1 (color online) (a) Scanning electron microscope images of the SRR dimer and its dimension. Gate voltage can be applied to each individual SRR to locally deplete the 2DEG underneath and guide the quantum Hall edge channel into the corresponding SRR gap. The two "cross-in-square" symbols represent the ohmic contacts. (b) Calculated electric field intensity (normalized to its maximal value) of the SRR dimer (without 2DEG), with the field probe positioned in one of the SRR gaps. (c) Color-coded in-plane electric field distribution of two resonant modes: the symmetric and the antisymmetric modes. (d) Calculated absorption spectra of the CR-SRR dimer coupled system. The fitted polariton branches corresponding to the bright symmetric mode are overlaid as red curves.

FIG. 2 (color online) Photocurrent spectra under (a) top-SRR gating ($V_{\text{Top}} < 0$), and (b) bottom-SRR gating ($V_{\text{Bot}} < 0$). USC to both symmetric and antisymmetric modes are detected, with the fitted polariton branches are overlaid as red and magenta curves, respectively. (c), (d) Configurations of edge channels under gate conditions in (a) and (b), respectively. Here, we denote the electrochemical potentials of the left (L) and right (R) contacts as $\mu_L$ and $\mu_R$, respectively; at zero source–drain bias, $\mu_L = \mu_R$. Since different geometrical edge channels enter the field-enhancement region, the photocurrent also exhibits chirality with respect to both the polarity of magnetic field and the "selected SRR".

FIG. 3 (color online) (a) Scanning electron microscope images of the SRR chain and its dimension. The two SRRs enclosed by the blue dashed rectangle constitute one unit cell. The two "cross-in-square" symbols represent the ohmic contacts. (b) Calculated electric field intensity (normalized to the maximal values) of the topological SRR chain (without 2DEG), evaluated at probe positions placed in the edge and bulk SRR gaps. (c), (d) Color-coded in-plane electric field distribution of the bulk modes and the topological edge mode.

FIG. 4 (color online) (a) Schematic illustration of the topological SRR chain device. A negative gate voltage is applied to either the edge ($V_{\text{Edge}}$) or the bulk ($V_{\text{Bulk}}$) SRR to detect the USC in different positions. Photocurrent spectra were measured at $V_{\text{SD}} = 0$, for which $\mu_L = \mu_R$. (b) Simulated absorption spectra of the CR-SRR chain coupled system as a function of magnetic field. (c), (d) Photocurrent spectra under bulk-SRR gating ($V_{\text{Bulk}} < 0$), and edge-SRR gating ($V_{\text{Edge}} < 0$), respectively. The fitted polariton branches associated with the bulk modes are shown as red and magenta solid curves, while the polariton branches associated with the topological edge mode are shown as red dashed curves.

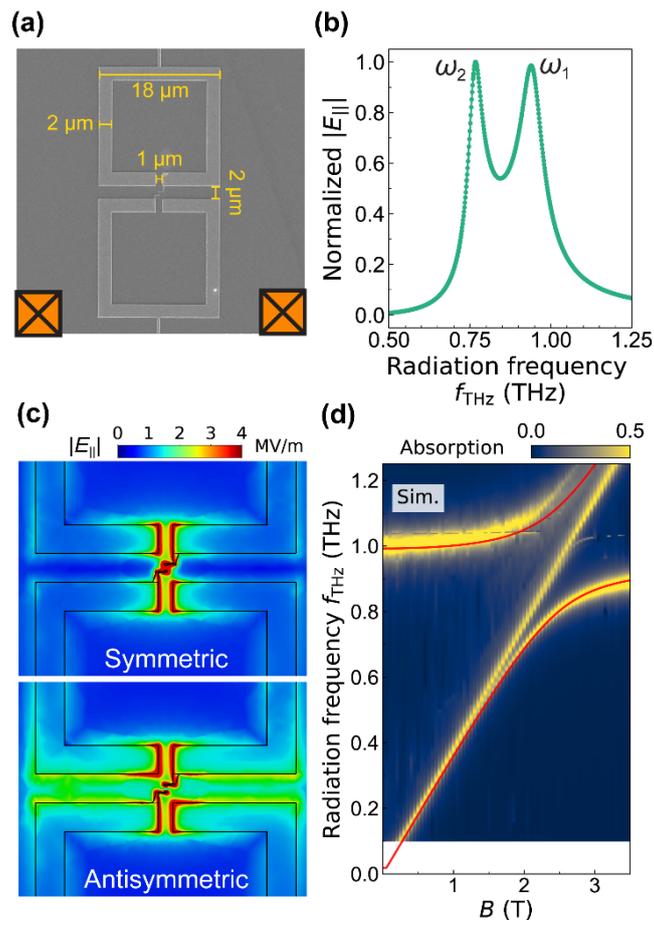

FIG. 1 HUANG, et al.

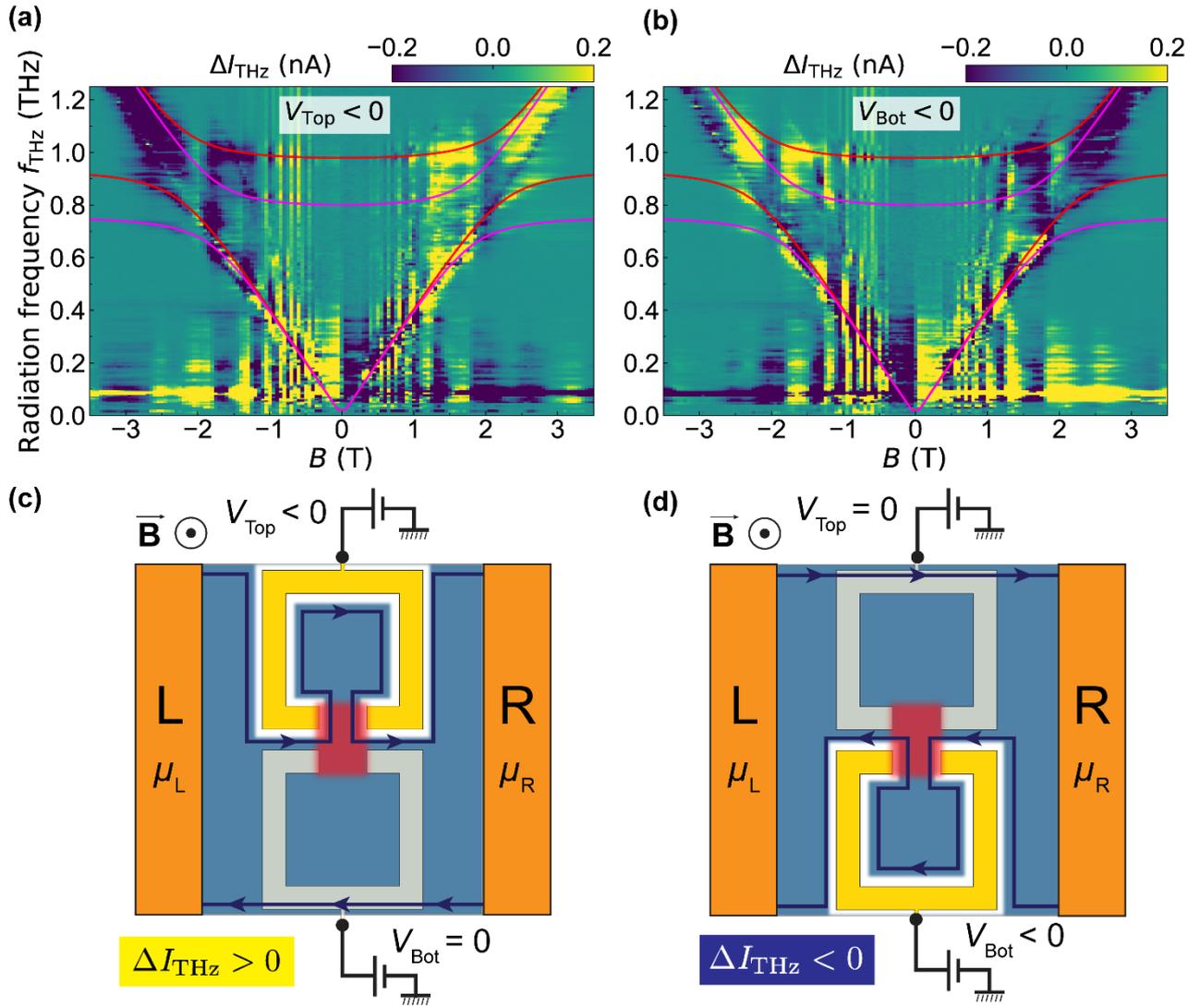

FIG. 2 HUANG, et al.

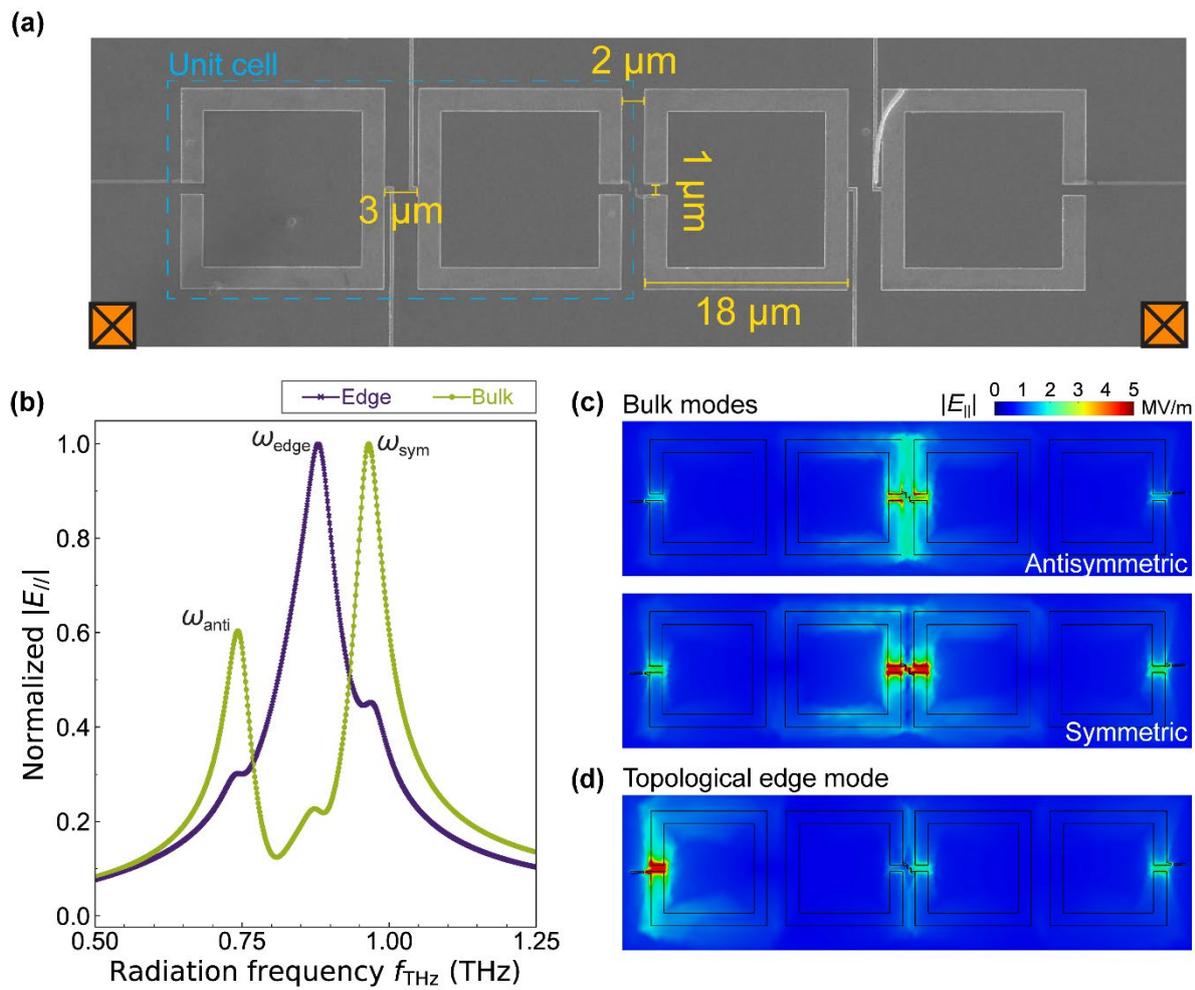

FIG. 3 HUANG, et al

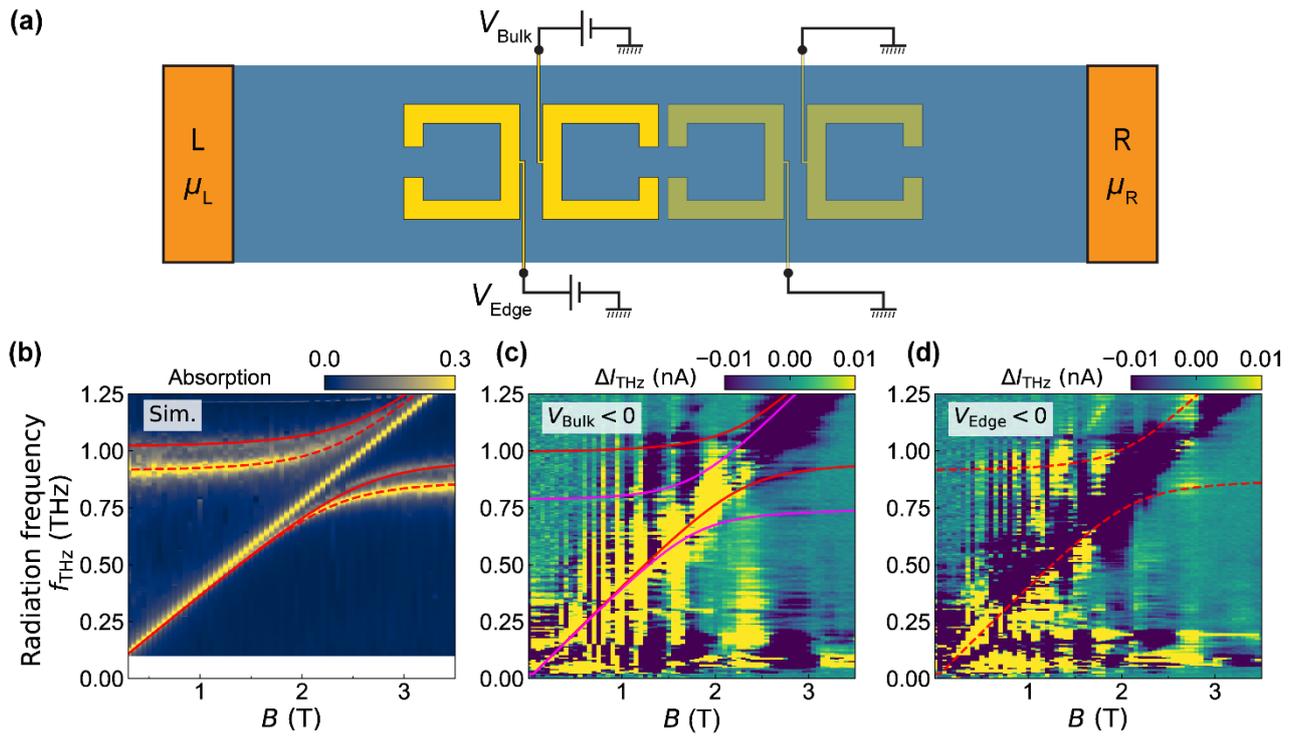
20FIG. 4 HUANG, et al.

**Supplementary Materials: Ultrastrong light–matter coupling in near-field coupled split-ring resonators revealed by photocurrent spectroscopy**

Jing Huang,[1] Jinkwan Kwoen,[2] Yasuhiko Arakawa,[2] Kazuhiko Hirakawa[1] and Kazuyuki Kuroyama,[1]

[1]*Institute of Industrial Science, University of Tokyo, 4-6-1 Komaba, Meguro-ku, Tokyo 153-8505, Japan*

[2]*Institute for Nano Quantum Information Electronics, University of Tokyo, 4-6-1 Komaba, Meguro-ku, Tokyo 153-8505, Japan*

**This supplementary material file includes:**

1. FE simulation for different SRR architectures
2. Electric field polarization of asymmetric mode and symmetric mode in SRR dimer
3. Semiclasscial simulation of coupled system
4. Optical absorption and photocurrent response of a single SRR and the underlying detection mechanism
5. Quantum mechanical model of Landau polariton with two cavity modes
6. Mismatch between spectra of SRR dimer coupled system and single-mode Hopfield model fits
7. Intracell coupling between SRRs: inductive coupling
8. Schematic of photocurrent measurement setup

# 1. FE simulation for different SRR architectures

We simulate the SRR structures presented in the main text using a commercial software (CST Studio Suite). For simplicity, the connection electrodes to each SRR were omitted in the simulations. The metallic structures were modeled using the predefined material parameters for gold in the software library, including its conductivity and loss. The SRRs are sandwiched between undoped loss-free GaAs substrate and vacuum.

## 1.1. Single SRR

For completeness, we first outline the simulation procedure used for a single SRR. The single SRR was enclosed in a cubic simulation domain whose lateral dimensions were set to at least twice the SRR size and treated as unit-cell boundaries in both the x- and y- directions. Floquet ports were defined on the top (vacuum side) and bottom (GaAs side) of the domain. A plane wave polarized along the SRR gap was launched from the bottom port to excite the SRR. The absorption spectrum of the single SRR was calculated as

$$Abs = 1 - \sum_{I,j} |S_{ij}|^2.$$

## 1.2. Near-field coupled SRR

In contrast, as for the SRR dimer and topological SRR chain, the asymmetric mode and the topological edge mode cannot be efficiently excited by the above scheme. To simulate their resonant behaviors, we replaced the plane-wave excitation with a localized current source placed near the gap of one SRR [1], and defined several electric-field probes in the gaps of neighboring SRRs. In this configuration, open boundary conditions were applied along all $x$-, $y$-, and $z$- directions. The resonance modes of the near-field coupled SRRs were then identified from the electric-field amplitude spectra measured at different probe positions.

## 2. Electric field polarization of asymmetric mode and symmetric mode in SRR dimer

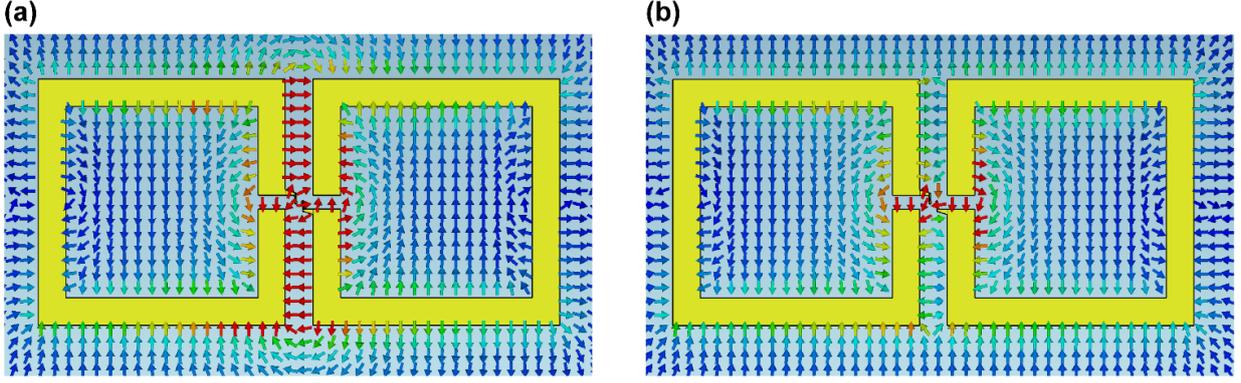

FIG. S1. Electric field polarizarion of (a) asymmetric mode and (b) symmetric mode in the SRR dimer. The opposite polarization in each SRR gap of the asymmertric mode results in a negligible electric dipole moment.

## 3. Semiclasscial simulation of coupled system

To obtain the optical transmission spectra of the coupled CR-SRRs system, we follow the approach described in Ref. [2]. The 2DEG is modelled by a gyrotropic layer below the GaAs surface with 50 nm thickness. The gyrotropic material can be regarded as a cold magnetized plasma described by Drude model. We bias the material a z-direction magnetic field, and the dielectric dispersion of the gyrotropic material is expressed by a complex permittivity matrix:

$$\varepsilon(\omega) = \begin{bmatrix} \varepsilon_1(\omega) & \varepsilon_2(\omega) & 0 \\ -\varepsilon_2(\omega) & \varepsilon_1(\omega) & 0 \\ 0 & 0 & \varepsilon_3(\omega) \end{bmatrix},$$

where

$$\varepsilon_1(\omega) = \varepsilon_\infty - \frac{\omega_p^2(\omega - i\nu_c)}{\omega(\omega - i\nu_c)^2 - \omega\omega_{CR}^2},$$

$$\varepsilon_2(\omega) = \frac{-i\omega_p^2 \omega_{CR}}{\omega(\omega - i\nu_c)^2 - \omega\omega_{CR}^2},$$

$$\varepsilon_3(\omega) = \varepsilon_\infty - \frac{\omega_p^2}{\omega(\omega - i\nu_c)}.$$

Here, $\varepsilon_\infty$ is the background dielectric constant, $\omega_p$ is the three-dimensional plasma frequency that depends on the electron density, $\omega_{CR}$ is the CR frequency as defined in the main text, and $\nu_c$ is the collision frequency. These parameters are adjusted to match the properties of the GaAs 2DEG.

In general, far-field transmission experiments require an array of identical hybrid systems [3-5] so that the excited near-field components can diffract into the far-field and be detected. To reproduce this condition in simulation, we model the CR-SRRs coupled system within a cubic simulation domain. The entire domain is treated as a unit cell in the x- and y-directions to emulate an infinite array of hybrid systems, while Floquet ports are defined at the top and bottom boundaries to calculate the transmission spectra, following the same configuration as in the single-SRR case.

## 4. Optical absorption and photocurrent response of a single SRR and the underlying detection mechanism

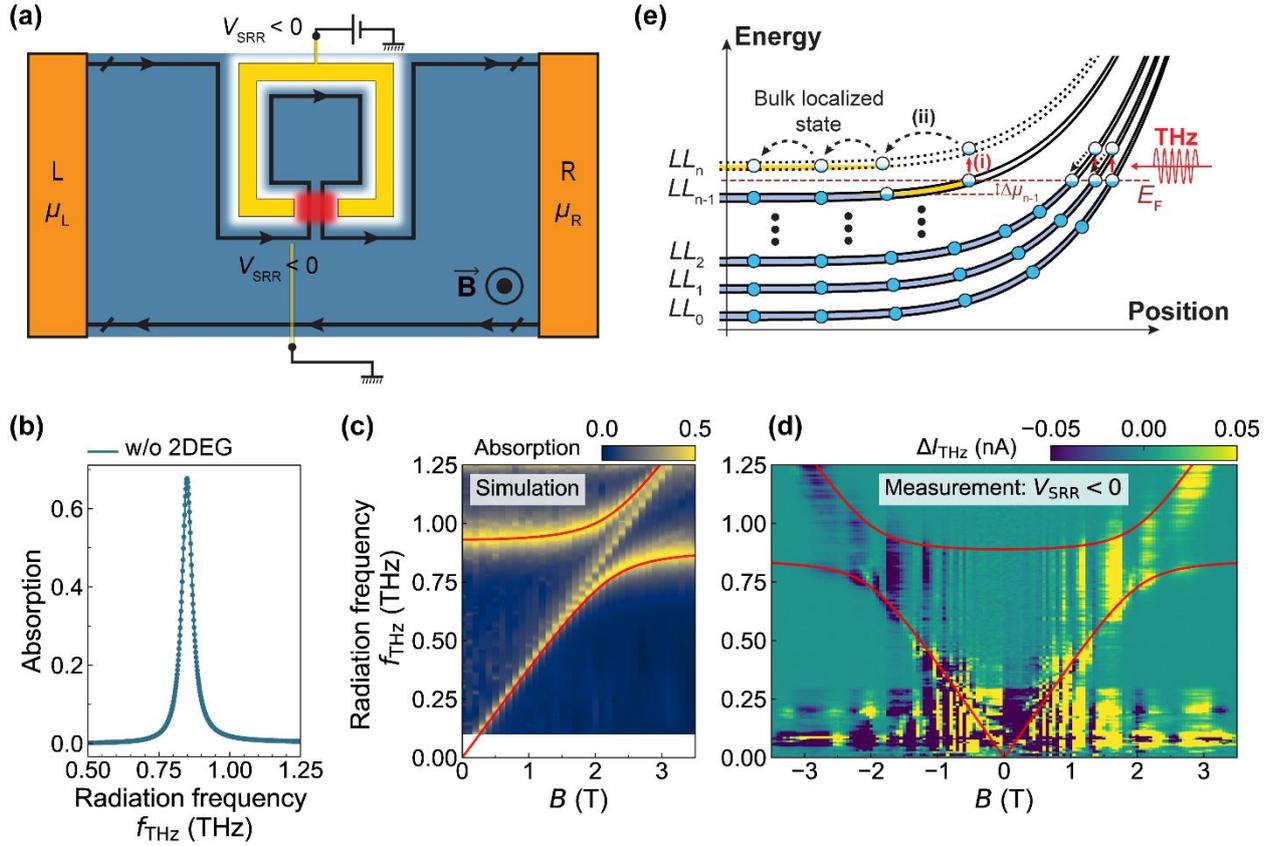

FIG. S2. (a) Schematic illustration of the single-SRR device. A negative gate voltage $V_{SRR}$ is applied to guide the edge channel into the SRR gap, where the terahertz cavity field induces a non-equilibrium electron distribution and generates photocurrent. (b) Calculated absorption spectrum of the single SRR (without 2DEG). (c) Simulated absorption spectra and (d) measured photocurrent spectra of the single SRR coupled to the CR as a function of magnetic field. The polariton branches are overlaid as red curves. (e) Schematic of the nonequilibrium electrochemical potential distribution in the edge channels induced by the generation and decay of polaritons.

As a reference, we describe the USC to a single SRR probed by optical absorption spectroscopy and photocurrent spectroscopy. A simplified schematic of the measurement setup is shown in Fig. S2(a). Although a side gate was fabricated near the SRR gap, no gate voltage is applied to it, and therefore it does not play any role in the present research. In Fig. S2(b), we report the calculated absorption spectrum of the single SRR. The resonance peak at $\omega_{SRR} = 0.85$ THz corresponds to the fundamental LC mode of the SRR. Figure S2(c) shows the simulated optical absorption spectra of the CR-SRR coupled system, which gives a normalized coupling ratio of $\Omega^{sim}/\omega_{SRR} = 0.15$ according to the fitting (red curves) with the single-mode Hopfield model. Figure S2(d) shows photocurrent spectra of the single SRR sample when a negative gate voltage is applied to the SRR. Typically, a similar anti-crossing feature is revealed by the chiral photocurrent response. The photocurrent spectra can be fitted by the Hopfield model with the same cavity frequency $\omega_{SRR}$ and coupling strength $\Omega^{exp} = \Omega^{sim}$ (red curves in Fig. S2(d)), which demonstrates the consistency between optical absorption measurement and photocurrent spectroscopy in the detection of USC.

The detection mechanism of the photocurrent spectroscopy is based on nonequilibrium electron distributions among multiple quantum Hall edge channels, which has been investigated in our previous studies [6-8]. Figure S2(e) illustrates the general case with a bulk filling factor $v$ = n (n = 0, 1, 2, ⋯), where the Fermi energy $E_F$ lies between the $LL_{n-1}$ and $LL_n$ ($LL_N$: the $N^{th}$ Landau level). Under terahertz irradiation, Landau polaritons are generated in the SRR gap and subsequently decay into their matter component, namely edge electron–hole pairs [process (i) in Fig. S2(e)] [7]. A fraction of the excited electrons in the innermost edge channel ($LL_{n-1}$) are then scattered to the bulk localized states by electron-phonon interaction or electron-electron interaction [process (ii) in Fig. S2(e)] and remain for a long lifetime [9-10]. This extraction of electrons shifts the electrochemical potential of $LL_{n-1}$ by $\Delta\mu_{n-1}$, giving rise to the detected photocurrent response. Similar electron–hole pair generation and relaxation processes may also occur between the outer LLs; however, since the excited electrons accumulate within the conductive edge channels, no photocurrent is produced in those cases. By using the Landauer-Büttiker formula and following the same analysis in [7], the photocurrent can be expressed as

$$\Delta I_{THz}^{v=n}(B) = sgn(B) \cdot \frac{2e}{h} \Delta\mu_{n-1}. \tag{1}$$

This equation reproduces the detected chiral photocurrent with respect to magnetic field, and explains why the photocurrent is more pronounced in the specific magnetic field region where edge-transport dominates the electron system.

## 5. Quantum mechanical model of Landau polariton with two cavity modes

We follow the general model developed in Ref. [11] and derive below the multi-mode Hopfield model used in the main text. The Hamiltonian describing the interaction between the inter-Landau-level transition and the electromagnetic field is given by

$$\mathcal{H} = \mathcal{H}_{cav} + \sum_{j=1}^{N}\hbar\omega_{CR}c_j^\dagger c_j + i\sqrt{\frac{\hbar\omega_{CR}e^2}{m^*}}\sum_{j=1}^{N}[c_j^\dagger A_-(\boldsymbol{r}_j) - c_j A_+(\boldsymbol{r}_j)] + \frac{e^2}{m^*}\sum_{j=1}^{N}A_-(\boldsymbol{r}_j)A_+(\boldsymbol{r}_j),$$

where $\mathcal{H}_{cav}$ is the bare cavity Hamiltonian, $N$ is the number of electrons, $c_j^\dagger$ ($c_j$) is the the bosonic creation (annihilation) operator of the local inter-Landau level excitation, and $A_\pm(\boldsymbol{r})$ are the non-Hermitian vector potential. The cavity vector potential with two modes can be expressed as

$$A_-(\boldsymbol{r}) = \sum_{\nu=1,2}\sqrt{\frac{\hbar}{2\varepsilon_0\varepsilon_r(\boldsymbol{r})\omega_\nu V_\nu}}f_\nu(\boldsymbol{r})(a_\nu^\dagger + a_\nu),$$

$$A_+(\boldsymbol{r}) = \sum_{\nu=1,2}\sqrt{\frac{\hbar}{2\varepsilon_0\varepsilon_r(\boldsymbol{r})\omega_\nu V_\nu}}f_\nu^*(\boldsymbol{r})(a_\nu^\dagger + a_\nu),$$

$$f_\nu(\boldsymbol{r}) = \frac{f_{\nu,x}(\boldsymbol{r}) - if_{\nu,y}(\boldsymbol{r})}{\sqrt{2}},$$

Here, $\varepsilon_r$ is the background dielectric constant, $\omega_\nu$, $V_\nu$, $a_\nu^\dagger$ ($a_\nu$) are the resonant frequency, cavity mode volume, and creation (annihilation) operator for mode $\nu$, respectively. $f_\nu$ is defined to be the normalized mode profile for mode $\nu$, while $f_{\nu,x}$ and $f_{\nu,y}$ are the normalized mode profiles for x- and y-polarization. In general, the two different cavity modes are orthogonal over the full space $V$:

$$\int_V f_\nu(\boldsymbol{r})f_\mu^*(\boldsymbol{r})d^3r = \delta_{\nu\mu}V_\nu,$$

However, when limited in a subspace, such as the 2DEG plane $S_{2DEG}$, this orthogonality does not hold anymore:

$$\sum_{j=1}^{N}f_1(\boldsymbol{r}_j)f_2^*(\boldsymbol{r}_j) = N\int_{S_{2DEG}}f_1(z_{2DEG},\boldsymbol{r}_\parallel)f_2^*(z_{2DEG},\boldsymbol{r}_\parallel)d^2\boldsymbol{r}_\parallel \neq 0.$$

Since we consider only the cavity field in the 2D plane, we can safely replace the full mode profiel $f_\nu(\boldsymbol{r})$ with an in-plane mode profile $f_\nu(z_{2DEG},\boldsymbol{r}_\parallel)$, and expand it in terms of an arbitrary orthonormal basis $\{\phi_1(\boldsymbol{r}_\parallel),\phi_2(\boldsymbol{r}_\parallel)\}$. Here, we choose the basis such that

$$f_1(z_{2DEG},\boldsymbol{r}_\parallel) = \alpha_{1,1}\phi_1(\boldsymbol{r}_\parallel),$$

$$f_2(z_{2DEG},\boldsymbol{r}_\parallel) = \alpha_{2,1}\phi_1(\boldsymbol{r}_\parallel) + \alpha_{2,2}\phi_2(\boldsymbol{r}_\parallel).$$

The sum of local inter-Landau-level transition can be simplified by introducing the collective bosonic matter operator:

$$b_\mu^\dagger = \frac{1}{\sqrt{2N}} \sum_{j=1}^{N} \phi_\mu(r_j) c_j^\dagger.$$

With the above definition of the collective matter operator, then the matter excitation mode $\mu = 1$ only interaction with cavity mode $\nu = 1$, while matter mode $\mu = 2$ interact with both of the two cavity modes $\nu = 1, 2$. Furthermore, we assume that the full mode profile in more dominant in for one polarization, e.g., the x-polarization, which is usually the case for real SRR. Under this approximation, the spatial mode profile becomes real function: $f_\nu(r) = f_\nu^*(r)$ and the system Hamiltonian becomes

$$\mathcal{H} = \sum_{\nu=1,2} \hbar\omega_\nu a_\nu^\dagger a_\nu + \sum_{\mu=1,2} \hbar\omega_{CR} b_\mu^\dagger b_\mu$$

$$+ i\hbar\Omega_{11}(a_1^\dagger + a_1)(b_1^\dagger - b_1) + i\hbar\Omega_{21}(a_2^\dagger + a_2)(b_1^\dagger - b_1) + i\hbar\Omega_{22}(a_2^\dagger + a_2)(b_2^\dagger - b_2)$$

$$+ \hbar\frac{\Omega_{11}^2}{\omega_{CR}}(a_1^\dagger + a_1)^2 + \hbar\frac{\Omega_{21}^2 + \Omega_{22}^2}{\omega_{CR}}(a_2^\dagger + a_2)^2 + \hbar\frac{2\Omega_{11}\Omega_{21}}{\omega_{CR}}(a_1^\dagger + a_1)(a_2^\dagger + a_2),$$

And the coupling strengthes read as

$$\Omega_{11} = \sqrt{\frac{e^2 \omega_{CR} N}{m^* \varepsilon_0 \varepsilon_r \omega_1 V_1}} \alpha_{1,1},$$

$$\Omega_{21} = \sqrt{\frac{e^2 \omega_{CR} N}{m^* \varepsilon_0 \varepsilon_r \omega_2 V_2}} \alpha_{2,1}, \qquad \Omega_{22} = \sqrt{\frac{e^2 \omega_{CR} N}{m^* \varepsilon_0 \varepsilon_r \omega_2 V_2}} \alpha_{2,2}.$$

To quantify the the mode overlap, we define the following quantities:

$$\mathcal{F}_{11} = \int_{S_{2DEG}} f_1(z_{2DEG}, r_\parallel) f_1^*(z_{2DEG}, r_\parallel) d^2 r_\parallel = |\alpha_{1,1}|^2,$$

$$\mathcal{F}_{22} = \int_{S_{2DEG}} f_2(z_{2DEG}, r_\parallel) f_2^*(z_{2DEG}, r_\parallel) d^2 r_\parallel = |\alpha_{2,1}|^2 + |\alpha_{2,2}|^2,$$

$$\mathcal{F}_{21} = \int_{S_{2DEG}} f_2(z_{2DEG}, r_\parallel) f_1^*(z_{2DEG}, r_\parallel) d^2 r_\parallel = \alpha_{2,1} \alpha_{1,1}^*,$$

which allows us to define the renormalized mode volumes with respect to the basis $\{\phi_1(r_\parallel), \phi_2(r_\parallel)\}$:

$$\tilde{V}_1 = \frac{V_1}{\mathcal{F}_{11}}, \qquad \tilde{V}_2 = \frac{V_2}{\mathcal{F}_{22}},$$

and eventually the coupling strengthes are given by

$$\Omega_{11} = \sqrt{\frac{e^2 \omega_{CR} N}{m^* \varepsilon_0 \varepsilon_r \omega_1 \tilde{V}_1}},$$

$$\Omega_{21} = \sqrt{\frac{e^2 \omega_{CR} N}{m^* \varepsilon_0 \varepsilon_r \omega_2 \tilde{V}_2}} \eta_{21} = \Omega_2 \eta_{21}, \qquad \Omega_{22} = \sqrt{\frac{e^2 \omega_{CR} N}{m^* \varepsilon_0 \varepsilon_r \omega_2 \tilde{V}_2}} \sqrt{1 - \eta_{21}^2} = \Omega_2 \sqrt{1 - \eta_{21}^2},$$

with the **mode overlap coefficient $\eta_{21}$**

$$\eta_{21} = \frac{\mathcal{F}_{21}}{\sqrt{\mathcal{F}_{11} \mathcal{F}_{22}}} = \frac{\alpha_{21}}{\sqrt{|\alpha_{2,1}|^2 + |\alpha_{2,2}|^2}}, \qquad 0 \le \eta_{21} \le 1.$$

A value of $\eta_{21}$ close to unity (zero) corresponds to strong (weak) spatial overlap between the two cavity modes.

In our study, we first calculated the mode overlap coefficient $\eta_{21}$ using FE simulation. Generally speaking, the spatial distribution of electric field of mode $\nu$ can be written as:

$$\boldsymbol{E}_\nu(\boldsymbol{r}) = \sqrt{\frac{N_\nu \hbar \omega_\nu}{2 \varepsilon_0 \varepsilon_r V_\nu}} (f_{\nu,x}(\boldsymbol{r}), \ f_{\nu,y}(\boldsymbol{r}), \ 0),$$

where $N_\nu$ denotes the number of photons in mode $\nu$ and the z-polarization is neglected. Using this representation, we find that

$$\eta_{21} = \frac{\int_{S_{2DEG}} \boldsymbol{E}_2(z_{2DEG}, \boldsymbol{r}_\parallel) \boldsymbol{E}_1^*(z_{2DEG}, \boldsymbol{r}_\parallel) d^2 \boldsymbol{r}_\parallel}{\sqrt{\int_{S_{2DEG}} \boldsymbol{E}_1(z_{2DEG}, \boldsymbol{r}_\parallel) \boldsymbol{E}_1^*(z_{2DEG}, \boldsymbol{r}_\parallel) d^2 \boldsymbol{r}_\parallel \cdot \int_{S_{2DEG}} \boldsymbol{E}_2(z_{2DEG}, \boldsymbol{r}_\parallel) \boldsymbol{E}_2^*(z_{2DEG}, \boldsymbol{r}_\parallel) d^2 \boldsymbol{r}_\parallel}},$$

where the right-hand side of the equation can be computed directly in CST Studio Suite. In our situation, we obtain a mode overlap coefficient $\eta_{21} = 0.19$.

As shown in Fig. S3 in Sec. 6, by fitting the transmission spectra using the multi-mode Hopfielde model ($\omega_1 = \omega_{sym}, \omega_2 = \omega_{asym}$), the small mismatch is removed and the extracted normalized coupling strengthes are $\Omega_1^{sim}/\omega_1 = 0.17$, $\Omega_2^{sim}/\omega_2 = 0.19$, with the overlapping coefficient $\eta_{21} = 0.19$, sitting in the weak overlap regime. Similarly, as for the photocurrent spectra shown in Fig. S4 in Sec. 6, the multi-mode Hopfield models gives a better fit to the spectra, with the similar overlapping $\eta_{21} = 0.19$ and normalized coupling ratios $\Omega_1^{exp}/\omega_1 = 0.15, \Omega_2^{exp}/\omega_2 = 0.16$.

## 6. Mismatch between spectra of SRR dimer coupled system and single-mode Hopfield model fits

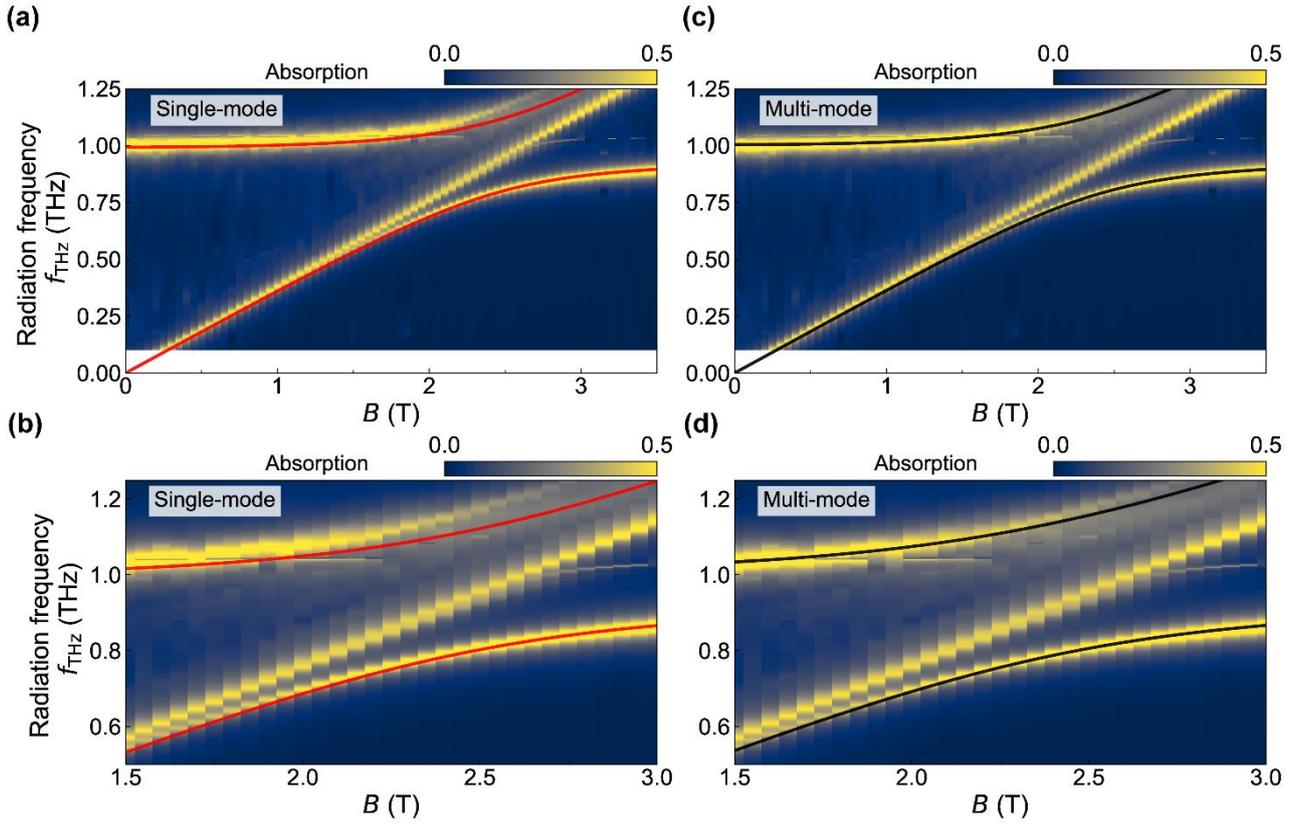

FIG. S3. Simulated transmission spectra of CR-SRR dimer coupled system overlaid by fitting using (a, b) single-mode Hopfield model (red curves) and (c, d) multi-mode Hopfield model (black curves) with $\eta = 0.19$. For simplicity, the other two polariton branches in the multi-mode model are not shown. The mismatch in single-mode fitting can be reduced by considering the multi-mode nature of SRR dimer, as discussed in Sec. 4.

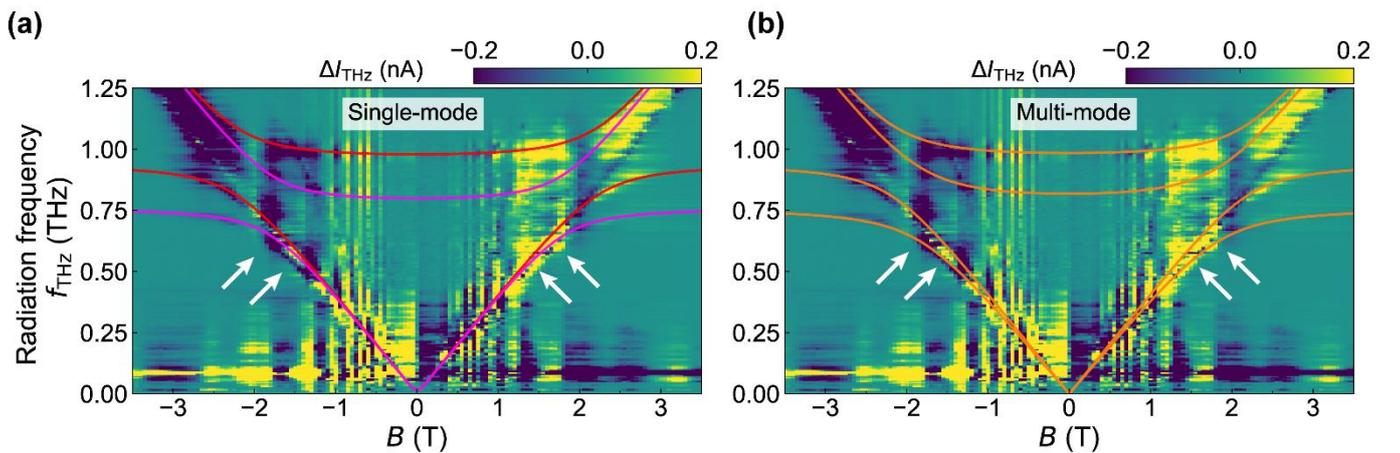

FIG. S4. Photocurrent spectra measured in SRR dimer fitted by (a) two separate single-mode Hopfield models (red and magenta curves) and (b) a multi-mode Hopfield model (orange curves) with $\eta = 0.19$. The multi-mode Hopfield model discussed in Sec. 5 yields better fit to the photocurrent spectra. As marked by the white arrows, considering a finite

mode overlapping allows the lowest polariton branch to reproduce the photocurrent spectra more accurately.

## 7. Intracell coupling between SRRs: inductive coupling

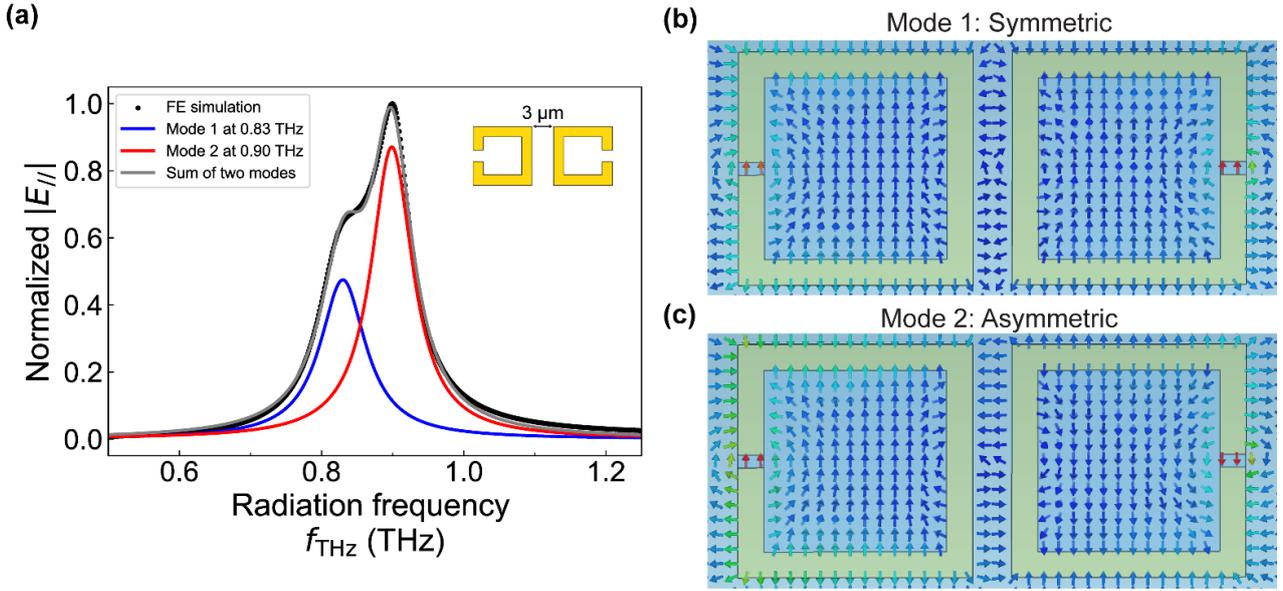

FIG. S5. (a) Simulated transmission spectra of SRR dimer with different orientation, where two SRR gaps are sitting far away from each other. (b, c) Electric field polarizarion of two modes in the inductive SRR dimer.

To confirm that the edge SRRs in the topological SRR chain are not isolated, we calculate the resonance spectrum of an inductively coupled SRR dimer, where the two SRR gaps are placed far apart from each other, as shown in the inset of Fig. S5(a). In this configuration, the coupling between SRRs is dominated by the effective magnetic dipole-dipole interaction and is therefore referred to as *inductive coupling*. In the FE simulation, the center-to-center distance between the two SRRs is set equal to that between the edge and bulk SRRs in the topological SRR chain so that we can extract the intracell coupling in our SRR chain. The appearance of mode splitting in the resonance spectrum indicates a finite coupling strength between the SRRs. The spectrum is fitted using two Lorentzian functions, from which the peak position are extracted as 0.83 THz and 0.90 THz, respectively. The electric-field polarization patterns of the two modes are shown in Figs. S5(b) and S5(c).

From the mode splitting, we estimate the intracell coupling to be 0.07 THz. As for the intercell coupling, since the separation of two bulk SRRs is identical to the SRR dimer device, the intercell coupling can be estimated to be 0.18. Therefore, the resulting ratio of the intracell coupling to the intercell coupling is 0.07/0.18=0.38<1. Additionally, the localization length can be estimated to be

$$l_{loc} = \frac{42\ \mu m}{ln(1/0.38)} = 100\ \mu m$$

where 42 um corresponds to the length of the unit cell.

## 8. Schematic of photocurrent measurement setup

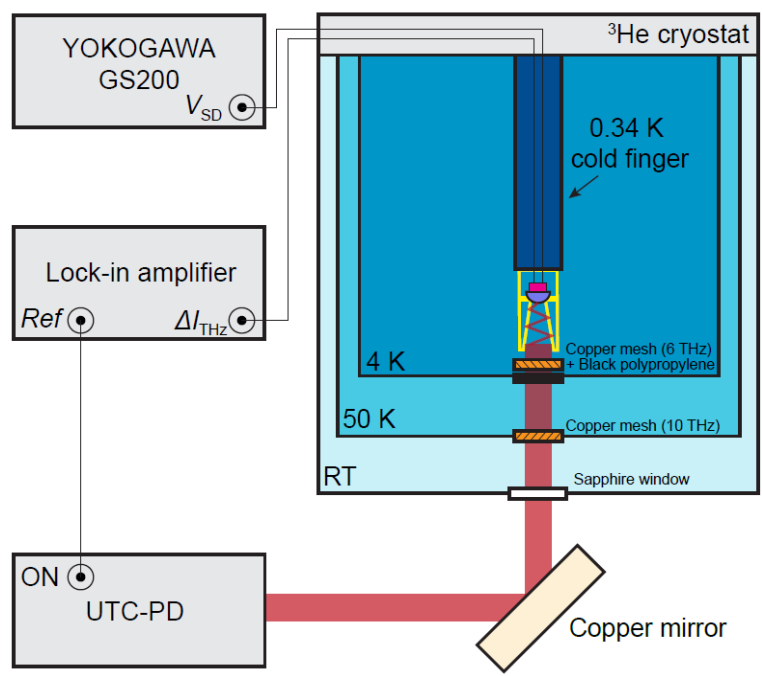

FIG. S6. Measurement circuit and optical path of photocurrent spectroscopy.